\documentclass{aa}  
\usepackage{graphicx}
\usepackage{txfonts}
\begin{document} 

   \title{On the mass mismatch between simulations and weak-lensing measurements}

   \author{J. Svensmark
           \inst{1}
           \and
           D. Martizzi
           \inst{1,2}\and
           A. Agnello
           \inst{1}\fnmsep\thanks{ORCID 0000-0001-9775-0331}
           }

   \institute{$^1$DARK, Niels Bohr Institute, University of Copenhagen,
              Lyngbyvej 2, 4. sal, 2100 Copenhagen \O, Denmark \\
              $^2$Department of Astronomy and Astrophysics, University of California, Santa Cruz, CA 95064, USA \\
              \email{jacob.svensmark@nbi.ku.dk , davide.martizzi@nbi.ku.dk , adriano.agnello@nbi.ku.dk}}


  \abstract
   {The recently discovered discrepancy between galaxy mass measurements from weak lensing and predictions from abundance matching questions our understanding of cosmology, or of the galaxy-halo connection, or of both.}
   {We re-examined this tension by considering, as models, different cosmological simulations in the Illustris suite. }
   {We produced \textit{excess profiles} $R\Delta\Sigma$ from subhalo snapshots at different redshifts in Illustris-1 and IllustrisTNG (TNG100, TNG300) simulations, enabling a direct comparison with weak-lensing measurements. We separate the individual contributions of stars, dark matter and gas within $\approx1$~Mpc (comoving length), beyond which correlated two-halo terms dominate.}
   {The mismatch between measurements and predictions is more severe than in previous studies: $R\Delta\Sigma$ profiles from IllustrisTNG are $\approx2$ times higher than the measured ones. Contrary to abundance matching results, the mismatch is mostly unchanged with increasing redshifts. The contribution of gas to the $R\Delta\Sigma$ profiles is $5-10\%$ over the scales dominated by one-halo terms.}
   {Different procedures to link stellar and halo masses (abundance matching, cosmological simulations) are still significantly discrepant with weak lensing measurements, but their trends are different. Therefore, the change in cosmological parameters advocated through abundance-matching arguments may not resolve this tension. Also, current criteria to select isolated massive galaxies in simulations are susceptible to resolution issues and may not correspond to observational criteria. The (currently subdominant) contribution of gas is non-negligible, and even if the major discrepancy within stellar and halo masses is resolved, it will be an appreciable source of systematics in the LSST era, when uncertainties on the $R\Delta\Sigma$ profiles are expected to be $\approx10$ times smaller.}


   \maketitle

\section{Introduction}

Our understanding  of cosmology is tied to that of the galaxy-halo connection. Departures from concordance cosmology, if any, should occur at scales $\lesssim 2$~Mpc, where however the mass budget and density profiles are also influenced by baryonic physics \citep[e.g.][]{par19,hil18}. Recent measurements from weak galaxy-galaxy lensing (GGL, from \citealt{leauthaud2017}, hereafter L17) resulted in mass profiles that cannot be reconciled with abundance-matching predictions based on clustering measurements \citep{saito16}, with a statistically significant mismatch. This mismatch has been confirmed independently and is seemingly robust against modelling and inference on the weak-lensing signal \citep{son19} and against different procedures to construct stellar-to-halo relations \citep{rod16,beh18}. This 
would then question our general understanding of cosmology or of galaxy formation, or both.

Here, we use different simulations from the Illustris suite to compare the mass profiles of massive galaxies to the measured GGL profiles of L17. We consider different redshift slices and multiple criteria to select isolated massive galaxies, including those by L17. We find that the mismatch has a different behaviour with redshift than what was found based on abundance-matching relations, and that galaxies satisfying the isolation criteria of L17 may actually not be isolated. We also find that such criteria are based on properties of simulated galaxies that are not entirely converged as a function of numerical resolution, which causes the GGL signal inferred from the selected galaxies to vary significantly from simulation to simulation. This paper is organised as follows: in Section~2 we recall the simulation suites used in this work and their main properties; the construction of weak-lensing profiles is described in Section~3; results are given in Section~4, and the implications are discussed in Section~5. For ease of comparison with the findings of L17, throughout this paper we use comoving distances to compute the weak-lensing mass profiles.

\begin{table*}
\caption{Overview of the available simulations of the Illustris \citep{vog14} and IllustrisTNG suite \citep{pil18} used in this work, through the publicly released catalogs and snapshot data \citep[see][for a release description]{Nel15}. $M_{\mathrm{bar}}$ shows the baryonic particle mass, $M_{\mathrm{dm}}$ the dark matter particle mass, $L_{\mathrm{box}}$ the box side length and $N_{\mathrm{dm}}$ the number of dark matter particles in the box. Rows $N_{a<z<b}$ show the number of subhalos with properties as discussed in the main text that resides within each redshift range $a<z<b$. Note that this number is distributed onto multiple snapshots. The number of snapshots within each redshift range is indicated in the second column.}
\label{table:simulations}      
\centering          
\begin{tabular}{c c c c c l l l l }
\hline\hline       
& &  \multicolumn{3}{c}{TNG300} & \multicolumn{3}{c}{TNG100} & Illustris-1 \\
& & 1 & 2 & 3 & 1 & 2 & 3 & 1 \\
\hline
   $M_{\mathrm{bar}}$   & $[10^6\,\mathrm{M_{\odot}}]$  & 11    & 88    & 703   & 1.4   & 11.2  & 89.2  & 1.3 \\ 
   $M_{\mathrm{dm}}$    & $[10^6\,\mathrm{M_{\odot}}]$  & 59    & 470   & 3760  & 7.5   & 59.7  & 478   & 6.3 \\ 
   $L_{\mathrm{box}}$   & [cMpc]                        & 302.6 & 302.6 & 302.6 & 110.7 & 110.7 & 110.7 & 106.5 \\ 
   $N_{\mathrm{dm}}$    &                               & $2500^3$&$1250^3$&$625^3$&$1820^3$&$910^3$&$455^3$&$1820^3$ \\
   $N_{0.4<z<0.5}$      & 4 snapshots   & 7136 & 3995 & 1913 & 610   & 363 & 202 & 778 \\ 
   $N_{0.5<z<0.6}$      & 5 snapshots   & 8228 & 4498 & 2142 & 737  & 424 & 240 & 1019 \\ 
   $N_{0.6<z<0.7}$      & 3 snapshots   & 4556 & 2429  & 1142 & 404   & 230 & 128 & 629 \\
\hline                  
\end{tabular}
\end{table*}

\section{Data} 
The Illustris project \citep{vog14}, and its TNG incarnation \citep{2017MNRAS.465.3291W,pil18,2018MNRAS.475..648P,spr18,2018MNRAS.475..624N,2018MNRAS.477.1206N,2018MNRAS.480.5113M}, comprise a suite of cosmological $\Lambda$-CDM simulations with N-body treatment of dark matter dynamics, magneto-hydrodynamical treatment of baryon dynamics, radiative cooling, star formation, stellar and active galactic nuclei feedback. The TNG project constitutes a significant update of the physical models implemented in the orginal Illustris suite. The simulation data from the original Illustris project and TNG are now publicly available \citep{Nel15,2019ComAC...6....2N}. A halo catalogue, built through a friends-of-friends (FoF) algorithm, is available for each snapshot of each simulation, with a corresponding subhalo catalogue within each halo generated using the {\sc subfind} algorithm \citep{2005Natur.435..629S}.

\begin{figure*}
    \centering 
    \includegraphics[width=1.0\textwidth]{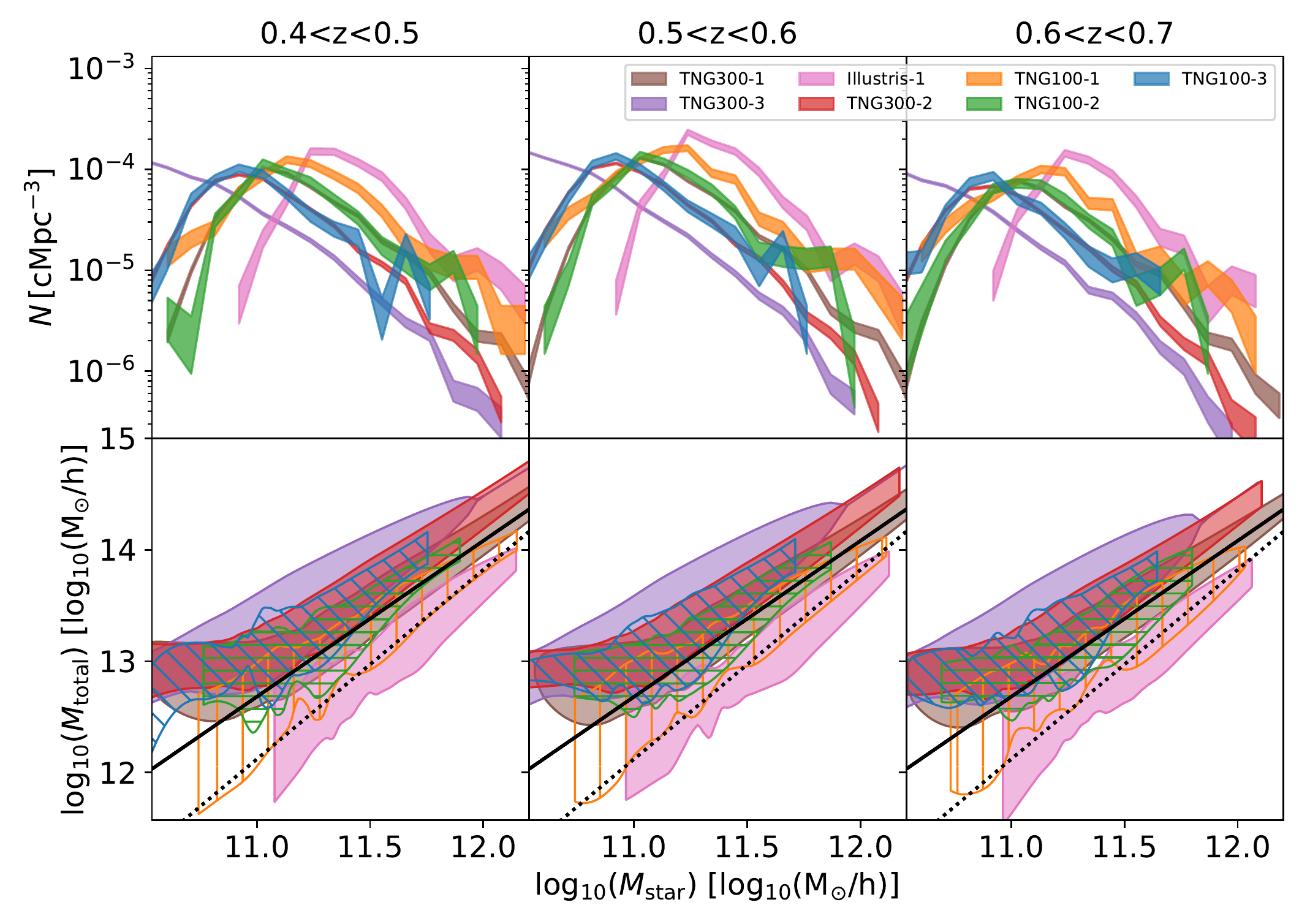}
    \caption{Stellar mass functions (top panels) and stellar-to-halo mass relations (bottom panels) from different Illustris-1/-TNG simulaitons, binned in the same redshift ranges as before. The stellar mass functions change significantly among simulations with different resolution, but (except for the poorest-resolution versions) they all agree around $M_{\star}\approx1.1\times10^{11}M_{\odot},$ i.e. only at the lower limit of our weak-lensing selection. The solid (resp. dotted) black line results from fits to strong (resp. weak) lensing data \citep{son18,son19}. }
  \label{fig:mstar}
\end{figure*}

The main properties of each simulation analysed in this work are summarized in Table~1. The public release provides simulation data for two box sizes of volume $110.7\,\mathrm{Mpc}^3$ and $302.6\,\mathrm{Mpc}^3$, which have been named TNG100 and TNG300, respectively. Each of the above has been run at three different resolution, both including and excluding the baryonic component. Given the box sizes and the abundance of massive galaxies, these are also the most suited to our study. 

\begin{figure*}
    \centering 
    \includegraphics[width=1\textwidth]{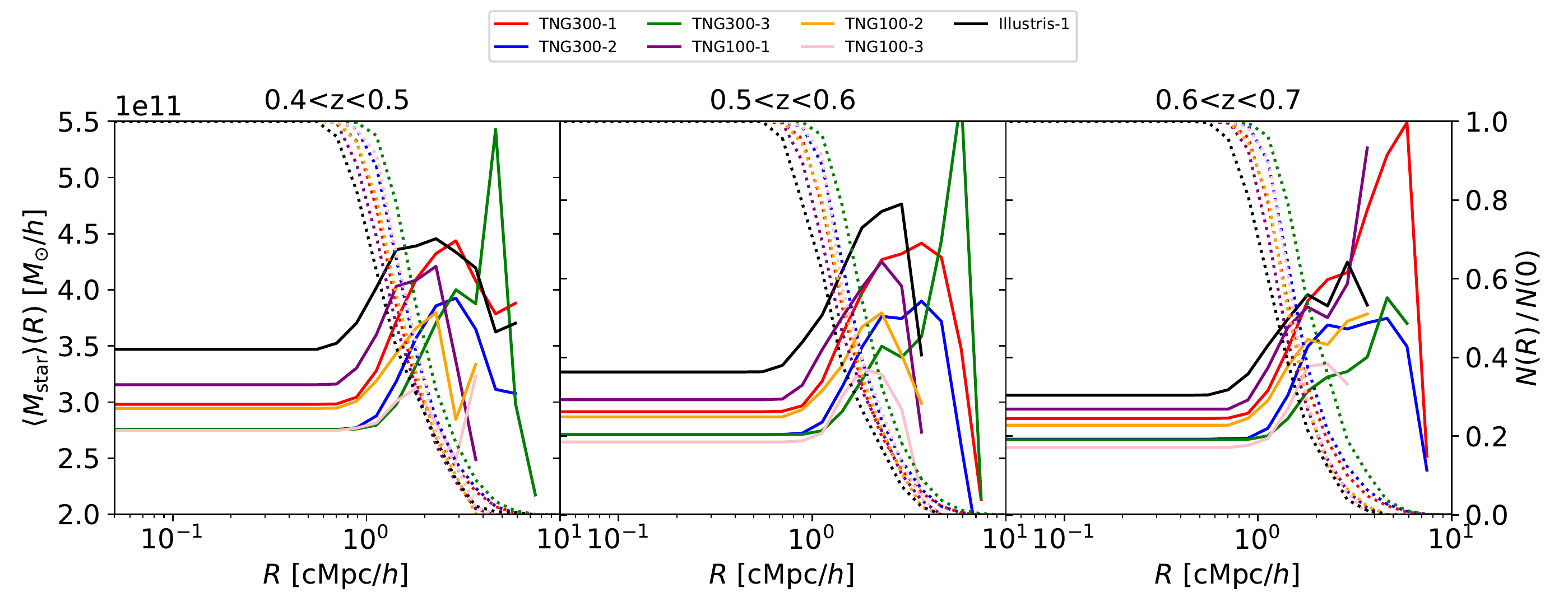}
    \caption{The contribution of acceptable subhalos to the GGL analysis in different Illustris-1/-TNG simulation boxes, in three redshift ranges corresponding to the same adopted by L17. 
    The dotted lines indicate the fraction of subhalos with non-zero $R\Delta\Sigma$ signal at different radii. The full lines show the average stellar mass $\langle M_{\star}\rangle(R)$ contributed by accepted subhalos at different locations.
    }
  \label{fig:diagnostics}
\end{figure*}

\section{The surface mass density contrast profile}
Following L17, we select halos of massive and isolated galaxies. In particular, within each simulation snapshot, we select subhalos with catalog stellar masses $M_{\star}>10^{11}\,M_{\odot}$ and a lower cutoff of the peak circular velocity $V_{\mathrm{max}}\geq351\,\mathrm{km/s}.$ 

The top panels of Figure~\ref{fig:mstar} show the stellar mass function of the velocity-selected subhalos in three redshift bins, $0.4<z<0.5$, $0.5<z<0.6$, and $0.6<z<0.7$, respectively. The stripes indicate the number of halos $N\pm\sqrt{N}$ within each bin in $M_{\mathrm{star}}$ assuming a Poisson distribution. The number density of galaxies with $M_{\star}\approx 10^{11}\,M_{\odot}$ is comparable across all simulations, and in multiple redshift bins, independently on box size and resolution. However, significant differences can be seen at higher stellar masses when resolution and box size are varied. This indicates that the simulations do not entirely converge for galaxies selected using our criteria, and that the effects of cosmic variance may also play a role in setting the observed differences. Nonetheless, it is encouraging to see that simulations with similar resolution (TNG300-1 and TNG100-2) yield similar galaxy stellar mass functions for our selection. 

The bottom panels of Figure~\ref{fig:mstar} show the relation between stellar mass and total subhalo mass. A \textit{Locally Estimated Scatterplot Smoothing} (LOESS) curve with variability bands was calculated for each simulation in the shown redshift bins, using a smoothing parameter of 0.35 \citep{msir,weisberg}. The plots show how with increasing resolution the TNG100 and TNG300 converge to similar relations. On the other hand, the original Illustris-1 simulation is systematically offset with respect to TNG100-1 and TNG300-1: for a fixed subhalo mass, Illustris-1 typically yields larger stellar masses than the TNG simulations. The TNG simulations also show a better match with the fit to strong lensing data provided by \cite{son18}, while the models of \citet{son19} on weak-lensng data show a better agreement with stellar-to-halo mass relations from Illustris-1. The offset between strong-lensing and weak-lensing relations is $\approx0.3$~dex (i.e. a factor $\approx2$), comparable to the scatter found in either relation.

We note that the \textit{integrated} densities of objects, following cuts in $V_{\mathrm{max}}$ (chosen by L17 as an isolation criterion), are the same across different simulations with comparable resolution, as reported in Table~1. However, there are two main issues with this choice. First, isolation criteria should be based on local densities of objects around given subhalos, which is not necessarily the case here. Second, a criterion based on $V_{\mathrm{max}}$ is susceptible to resolution in the simulations, and so may not even be a robust indicator of integrated density. Here, we have chosen to follow the isolation criteria by L17 for ease of comparison with their results, based on $z\approx0.5$ snapshots from Illustris-1.

For each subhalo within the selected range, the snapshot data of particles within its parent halo were extracted. A randomly oriented orthonormal basis was chosen and the parent halo particles were projected along its three directions. For each of the projections, the surface mass density $\Sigma (R)$ was calculated in logarithmically spaced radial bins, as well as the average surface density $\Sigma (<R)$ within circles of corresponding radii. This yields the surface mass density contrast profile
\begin{equation}
    \Delta \Sigma = \Sigma (R)-\Sigma (<R)
\end{equation}
which is the observable extracted from weak GGL measurements. This procedure was repeated for each accepted subhalo (satisfying the cuts on $M_{\star}$ and $V_{\mathrm{max}}$) and for different species of particles (stellar, dark matter, gas, black holes)\footnote{Tables of profiles for selected snapshots of all selected subhalos are publicly available upon request.}. The number of acceptable subhalos for each simulation box, in three redshift ranges, is summarized in Table~\ref{table:simulations}.

Quite expectedly, subhalos of different mass contribute to the GGL signal in different radial ranges. For this reason, we also recorded the fraction and average stellar mass of contributing subhalos in each radial bin. This is shown in figure~\ref{fig:diagnostics}, for different Illustris-1/-TNG simulations and in three redshift ranges matching the measurements of L17. The subhalo contributions are robust within $R\lesssim0.7$~Mpc. This is also comparable to the radius beyond which also correlated two-halo terms would start to dominate. The average stellar mass of contributing subhalos at $R\lesssim1$~Mpc is uniform with radius, but it changes across simulations as a result of the differing stellar-mass functions (shown in Fig.~1).

A possible source of concern would be the relation between shear, which is closer to what is actually measured in observations, and the $R\Delta\Sigma$ profiles. However, our criteria for isolation and the stacking over many subhalos ensures that the $R\Delta\Sigma$ is robustly determined, and a full ray-tracing simulation is not necessary. This is reinforced by the fact that our $R\Delta\Sigma$ profiles from Illustris-1 at $z\approx0.5$ are the same as obtained\footnote{In that work, only $z\approx0.5$ profiles from Illustris-1 were examined.} by L17.


\begin{figure*}
    \centering 
    \includegraphics[width=1.0\textwidth]{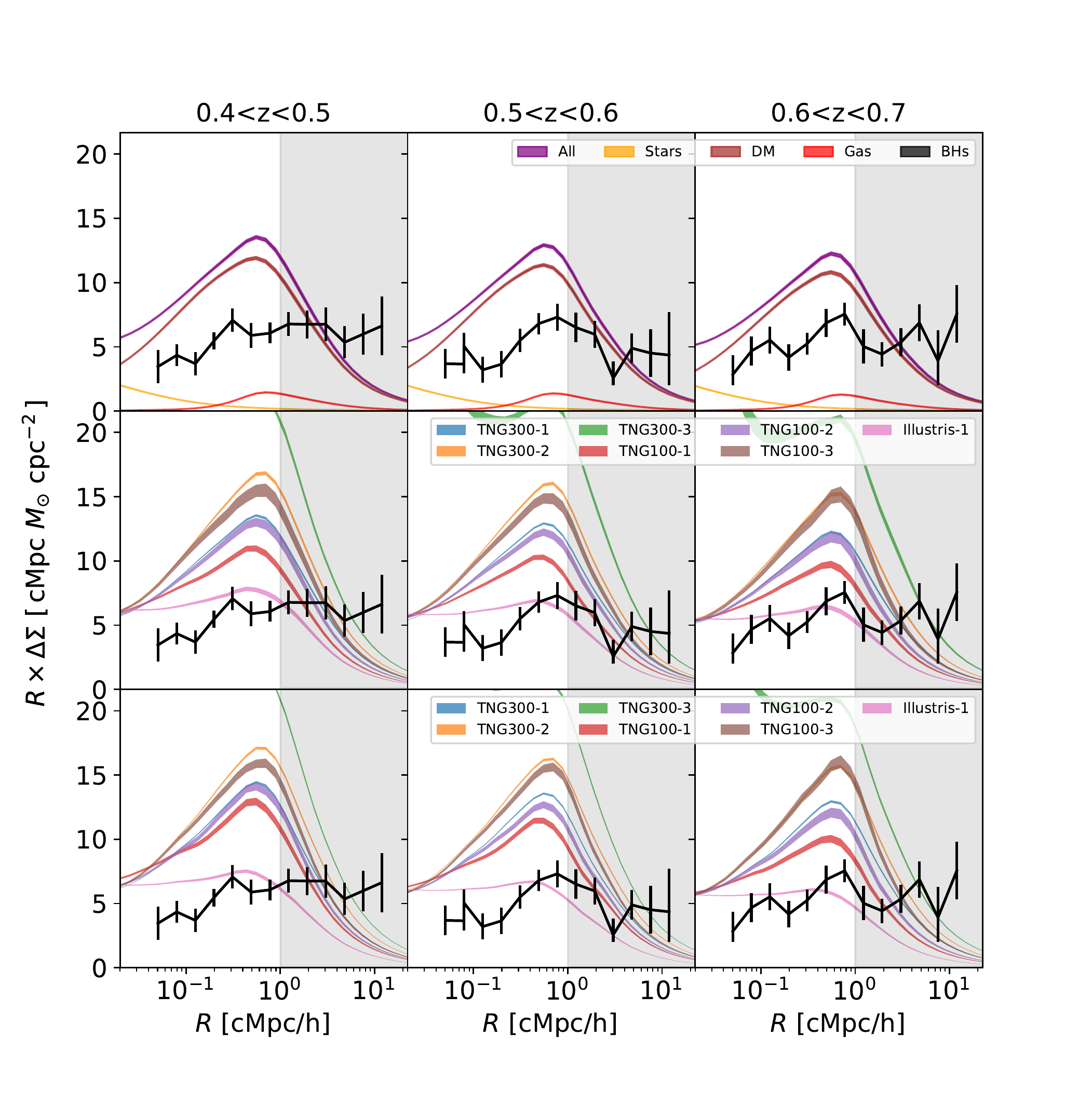}
    \caption{The GGL contrast profile $R\Delta\Sigma$ in three redshift ranges. \textit{Top:} Profiles from the TNG300-1 simulation, split into the four particle species (stars, dark matter, gas, black holes).
    That the black hole contribution is nearly zero and hardly visible in the figure.
    For each panel, the GGL profile measured by 
    L17 is shown as the black curve with errorbars. \textit{Middle:} The $R\Delta \Sigma$ profiles from different Illustris-1/-TNG box sizes and resolutions. In all panels, the width of the line is the 
    standard deviation of $R\Delta\Sigma$ across acceptable subhalos. \textit{Bottom:} Same as middle row, but retaining only subhalos whose projected stellar mass at $100\,\mathrm{kpc}$ is smaller than the catalog $M_{\mathrm{star}}.$ All length measures are in co-moving coordinates, as in L17.}
  \label{fig:zbins}
\end{figure*}

\begin{figure}
    \centering 
    \includegraphics[width=0.48\textwidth]{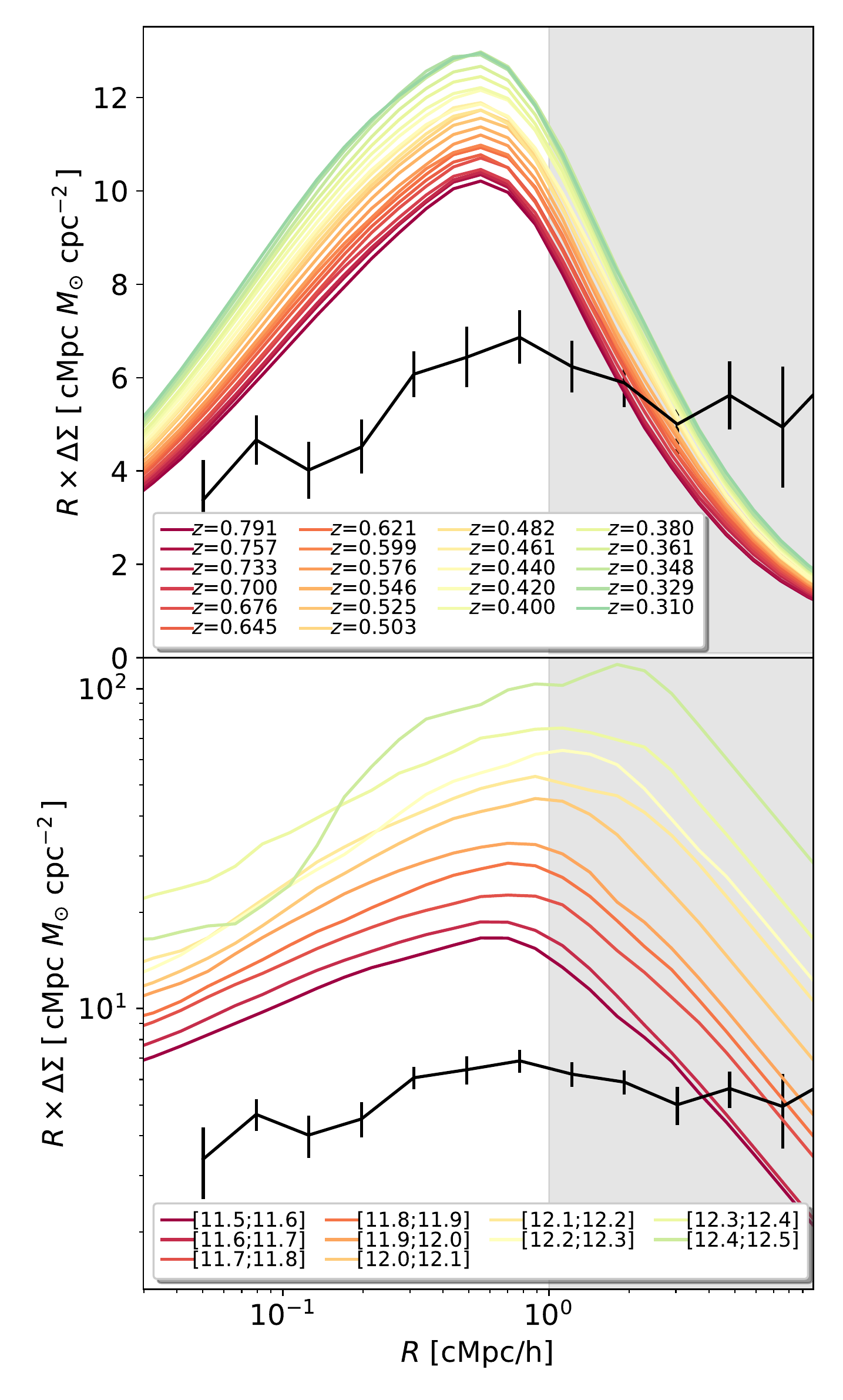}
    \caption{Top: Redshift evolution of the $R\Delta \Sigma$ profile for the TNG300-1 simulation, as represented by the coloured lines. The error-bars are the CMASS measurements as obtained from \cite{leauthaud2017}. Bottom: The $R\Delta \Sigma$ profile for the TNG-300 simulation at redshift $z=0.5,$ split in stellar masses within the range indicated by the legend. The ranges are $\log _{10}(M_{star})$, where $M_{star}$ is in $M_{\odot}/h$.}
  \label{fig:z_and_m_evolution}
\end{figure}

\section{Results}

From the subhalo tables generated in the previous section and for each particle type (dark matter, stars, gas and black holes), the excess profiles were averaged 
within 3 redshift ranges: $[0.4<z<0.5],$ $[0.5<z<0.6],$ and $[0.6<z<0.7].$ This was done for the sake of direct comparison with the measured GGL profiles by L17. The stacked profiles are shown in Fig. \ref{fig:zbins}, and the L17 GGL measurements are displayed with black error-bars. The top row displays the profile for each of the four particle types as well as the total, for the TNG300-1 simulation. The general tendency for the TNG300-1 excess profiles is to increase with decreasing redshift. 

The second row of Fig. \ref{fig:zbins} shows only the total excess profile, but for all of the simulations considered in this work. The same increase in the excess profiles with decreasing redshift can be seen for all of the simulations. Generally all simulations agree at low redshifts except for the lowest resolution `big box' simulation TNG300-3, as well as the older Illustris-1 result. With increasing redshift, the profiles spread out more, yet generally decrease towards the observed data. This is opposite to the models used by L17, based on abundance-matching through clustering \citep[by][]{saito16}, whose mismatch with measured profiles increased with increasing redshift.

The bottom row in Fig. \ref{fig:zbins} is analogous to the middle row, but with a more stringent isolation criterion on subhalos. In particular, we retained only subhalos whose mass in stellar particles within 100$\,$kpc was smaller than the subhalo $M_{\star}$ from the tabulated catalog. This removed $\approx2/3$ of the subhalos that satisfied the cuts of L17. The trend of the excess profiles, as well as of the mismatch with measured GGL profiles, is roughly the same.

Figure \ref{fig:z_and_m_evolution} shows the evolution the $R\Delta\Sigma$ excess profile of subhalos within the TNG300-1 simulation with redshift (top panel) and stellar mass (bottom panel). The redshift evolution is shown through 22 redshifts from the simulation snapshots between $z=0.791$ and $z=0.310$ for all the subhalo masses selected here. The mass evolution of the subhalos in the bottom figure is shown at $z=0.503$ for 20 $M_{\mathrm{star}}$ bins as shown in the legend, between $1\times 10^{11}\,M_{\odot}/h$ and $5\times 10^{12}\,M_{\odot}/h$. The excess profile is seen to depend strongly on mass, with increasing excess for increasing stellar mass.



\section{Discussion and Conclusions}

In this paper, we measure the GGL signal predicted by the Illustris and IllustrisTNG (TNG100 and TNG300) suite of cosmological hydrodynamical simulations. We select subhalos with stellar masses $M_{\star}>10^{11}\,M_{\odot}$ and a lower cutoff of the peak circular velocity $V_{\mathrm{max}}\geq351\,\mathrm{km/s}$, then  we further apply criteria for the isolation of halos, to investigate their impact on the GGL signal. L17 found a mismatch between GGL measurements and model predictions, considering abundance-matching models based on clustering, as well as the then available Illustris-1 suite (only showed for $z\approx0.5$ there).
We confirm the existence of a similar mismatch over the full suite of IllustrisTNG cosmological simulations, in different redshift ranges. However, contrary to the findings of L17 on Illustris-1, the mismatch is mostly unchanged with redshift.


Given the statistical uncertainties on the profiles, this mismatch is not only a byproduct of cosmic variance across different simulation boxes. However, this effect is still small compared to the mismatch with measured profiles. 

One of possible of sources of discrepancy between the GGL profiles measured from different simulations is the shape of the stellar mass function of the samples selected with our criteria. We find that the latter is influenced by both numerical resolution effects and cosmic variance. Applying a circular velocity cut allows us to construct galaxy samples with similar {\itshape integrated} number density in each simulation, but the samples have very different stellar mass functions. This issue is caused by the fact that stellar mass and circular velocity (i.e. the internal dynamics of dark matter halos) are properties that are not numerically converged at the resolutions reached by Illustris/IllustrisTNG. Differences in the stellar mass function yield differences in how massive galaxies are weighted against less massive galaxies when computing the stacked profiles, i.e. the GGL signal. Due to this effect, shallow tails at the high-mass end of stellar mass function produce a higher GGL signal. 

Another noteworthy effect is the GGL excess profile from gas particles. With the current mismatch, this $\approx5-10\%$ contribution to the GGL signal is still negligible, but it may become a significant source of systematics if the mismatch is resolved. This will hold especially in the regime of LSST measurements, where (by a simple scaling of footprint) the uncertainties in measured GGL profiles are expected to shrink by a factor $\approx10.$

These findings have important implications if the role of cosmological parameters is examined. The mismatch between simulations and observations may be resolved if for a fixed stellar mass near the peak of the $M_{*}/M_{total}$ ratio, the expected host halo mass were $\sim 0.3$~dex smaller than predicted by the IllustrisTNG simulations. Coincidentally, this is also the mismatch between strong-lensing and weak-lensing relations between stellar and halo masses \citep{son18,son19}.
As an alternative to the mismatch with models based on abundance matching by \citet{saito16}, L17 advocated for a lower value of the clustering amplitude $S_{8}\propto\sigma_{8}\sqrt{\Omega_{m}}$ with respect to \textit{Planck} measurements \citep[see][and refernces therein]{planck2018}. Similar claims had also been made in the past, considering the pairwise-velocity dispersions of galaxies and their clustering strength \citep{yan03,li09}. In fact, also the clustering of galaxies in IllustrisTNG \citep{spr18} is admittedly higher than current observational results. However, as shown by L17, a smaller value of $S_8$ would make the GGL mismatch decrease further with increasing redshift, and this may actually exacerbate the mismatch between measurements and profiles from the IllustrisTNG simulations. Moreover, the model GGL profiles examined in this work span length-scales where two-halo terms are negligible, so the discrepancy found by L17 may be linked more fundamentally to uncertainties in stellar masses and the construction of abundance-matching relations, rather than to the relation between clustering and cosmological parameters. 


\begin{acknowledgements}
      This project is funded by the Danish council for independent research under the project ``Fundamentals of Dark Matter Structures'', DFF - 6108-00470.AA was supported by a grant from VILLUM FONDEN (project number 16599). DM acknowledges support by the Danish National Research Foundation (DNRF132)
\end{acknowledgements}


\end{document}